\documentclass[10pt]{article}
\usepackage{graphicx}
\usepackage{latexsym}
\usepackage{amsfonts}
\usepackage{amscd}
\usepackage{amssymb}
\usepackage{amsthm}
\usepackage{amsmath}

\begin{document}
\title{On the effect of temperature on the reentrant condensation in polyelectrolyte-liposome
complexation}

\author{
S. Sennato $^{\S}$, D. Truzzolillo $^{\S}$ F. Bordi $^{\S}$, C.
Cametti  $^{\S}$
 \\
\small\it{$^{\S}$Dipartimento di Fisica,  Universita' di Roma ''La Sapienza''}\\
 \small\it{Piazzale A. Moro 5, I-00185 - Rome (Italy) and INFM CRS-SOFT, Unita' di Roma 1}}
\date{\today}
\maketitle

\begin{abstract}
Interactions of oppositely charged macroions in aqueous solution
give rise
 to intriguing aggregation phenomena, resulting in finite-size, long-lived
 clusters,
 characterized by a quite narrow size distribution. Particularly, the adsorption of
 highly charged linear polyelectrolytes on oppositely charged colloidal particles is strongly
 correlated and some short-range order arises from competing electrostatic
 interactions between like-charged polymer chains (repulsion) and between polymer chains and particle surface
 (attraction). The charge inversion observed for the polyelectrolyte-decorated primary particles
   has been recently explained in terms of correlated
  adsorption. In these systems, long-lived clusters of polyelectrolyte-decorated particles
  form in an interval of concentrations around the inversion point.
However, the mechanisms that drive the aggregation and
stabilize, at the different polymer/particle
 ratios, a well defined size of the aggregates are not completely understood. Nor is clear
  the role that the correlated adsorption plays in the aggregation, although the importance
  of 'patchy interactions' has been stressed as the possible source of attractive interactions between colloidal particles.
   Different models have been introduced to explain the formation of the observed
   finite-size cluster phase.
   However a central question
   still remains unanswered, i.e., whether the clusters are true equilibrium or metastable aggregates.
   To  elucidate this point, in this work, we have investigated the effect
   of the temperature on the formation of the clusters. We employed
   liposomes built up by DOTAP lipid interacting with a simple
   anionic polyion, sodium polyacrylate, over an extended
   concentration range below and over the isoelectric condition. Our
   results show that the aggregation process can be described by a
   thermally-activated mechanism.

\end{abstract}

\section{Introduction}

Charged colloidal particles in the presence of highly charged linear
polyelectrolytes self assemble in long-lived, finite-size mesoscopic
aggregates. Despite the increasing body of experimental and
theoretical work
\cite{Raspaud99,Wang00,Bordi04,Bordi06,Yaroslavov07,Yaroslavov07b,Radeva07,Volodkin07,Macdonald07,Bordi07c,Borkovec08},
the mechanisms that drive the aggregation and  stabilize the
aggregate size are not completely understood.

The strong and relatively long-ranged electrostatic interactions,
the screening due to the diffuse ionic layers surrounding the
particles, and the simultaneous presence of macroions different in
geometry and in charge density  (the linear polyelectrolytes and the
bulky particles), make these systems inherently complex. The
formation of the clusters is hence the result of a delicate balance
between attractions and repulsions characterized by different ranges
and also differing in their nature (mainly electrostatic
\cite{Volodkin07} but also hydrophobic \cite{Sennato08,
Yaroslavov07} interactions).

The adsorption of linear polyelectrolytes on the surface of the
oppositely charged particles occurs in a highly correlated manner.
The competing interactions between polymers and surface (attraction)
and between the polymers (repulsion) results in a locally ordered,
correlated adsorption of the chains
\cite{Mou95,Nguyen00,Dobrynin00,Grosberg02,Pericet04}. This
non-uniform distribution  gives rise to a non-uniform distribution
of the surface electric charge, with alternate patches where the
charge of the polymer or that of the bare particle surface are
locally in excess. This correlated adsorption is the cause of the
counterintuitive phenomenon of the 'overcharging', or charge
inversion, which occurs when more polyelectrolyte adsorbs than is
needed to neutralize the original charge of the surface, so that
 the net charge of the polymer-decorated surface inverts
its sign.

From the first observations of the destabilizing effect of
polyelectrolytes on colloidal suspensions, attractive 'bridging'
forces have been proposed as the cause of aggregation
\cite{LaMer63}. However, when a polyelectrolyte solution and a
suspension of oppositely charged colloidal particles  are mixed
together, the adsorption of the polymer on the particles and the
aggregation occur on two different timescales. Being favored by
the huge surface/volume ratio of the colloid, with the surface
'finely distributed' within the whole volume of the host phase,
the adsorption should be completed in a few seconds from the
mixing. Conversely, the aggregation process, due to the lower
diffusivity of the bulkier colloidal particles should occur on a
longer timescale. To this reasonable picture has been recently
given a solid experimental basis \cite{Volodkin07}.

As a consequence of the different timescales, the 'primary
particles' that are involved in the aggregation process are the
polyelectrolyte-decorated particles. This observation is not to
be overlooked, since it suggests a useful shift of the
perspective to guide the investigation of this phenomenology.
Rather than hypothesizing complex scenarios of
'polyelectrolyte-mediated' interactions, where the
polyelectrolytes, free in the solution among the particles,
drive the aggregation of the colloid by a mix of osmotic,
electrostatic screening and bridging effects, one has to deal
with a much simpler system of (non-uniformly) polymer-coated
particles.

There is much experimental evidence \cite{Mowald01,Henon03} as
well as theoretical arguments \cite{Dobrynin00, Dobrynin01,Li06}
in favor of a configuration of the adsorbed polyelectrolyte
chains that, although with several loops and possibly dangling
ends, lie quite flat on the particle surface, particularly if
the the charge densities on the surface and/or on the chains are
high \cite{Henon04}. Also these considerations suggest that
bridging mechanism should be excluded as the cause of
aggregation, except perhaps for very long polyelectrolytes and
very small particles.

The aggregation of charged colloids induced by oppositely
charged polyelectrolytes, and the stability of the resulting
clusters, appears mainly governed by electrostatic forces
\cite{Gregory73,Bordi07c,Borkovec08}. While repulsions derive
the electrostatic diffuse layer overlap, as described in the
classical Derjaguin-Landau-Verwey-Overbeek theory (DLVO),
attractions, besides the ubiquitous van der Waals dispersion
forces, include an additional non-DLVO attractive force,
originating from 'patch-charge' heterogeneities
\cite{Miklavic94,Khachatourian98,Velegol01}. In a broad sense,
this interaction can still be viewed as a sort of bridging,
since chain configurations span the inter-particle region to
gain substantial entropy \cite{Podgornik95}.

On the basis of the classical DLVO theory, one expects a stable
suspension far away from the isoelectric point (IP), and a
destabilization, with the formation of larger and larger aggregates
that eventually precipitate, close to the IP. In fact,  associated
to the progressive reduction of the net charge of the
polyelectrolyte-decorated particles, an increase of the cluster size
is observed. The largest size of the aggregates is observed close to
the isoelectric point, while any further addition of
polyelectrolytes promotes a progressive decrease of their size.
Eventually, when the particle surface is completely saturated by the
adsorbed polyelectrolyte and the overcharging has reached its
plateau value, the size of the particles in the suspension equals
again the size of the primary particles ('reentrant condensation')
plus a thin layer of adsorbed polymer \cite{Bordi04, Volodkin07,
Yaroslavov07b}. This reentrant condensation has been observed in a
variety of colloidal systems, ranging from polyelectrolyte-micelle
complexes \cite{Wang00}, latex particles \cite{Keren02}, dendrimers
\cite{Kabanov00}, phospholipid vesicles (liposomes) \cite{Radler98,
Bordi06,Volodkin07}, to 'hybrid niosomal' vesicles \cite{Sennato08}.
Although the adsorption patterns are further complicated by the
presence of short-range interactions specific to the different
systems, the similarities in the observed behavior for such
different systems strongly indicate that the overall phenomenology
is mainly governed by electrostatic interactions, arising from
double layer overlap (repulsion) and surface charge non-uniformity
(attraction).

DLVO theory does not account for these long-lived finite-size
aggregates. By itself, the theory only provides a description of the
stabilization/destabilization behavior. Once the colloid had been
destabilized, the size of the clusters evolve along dynamics that
depends on both particle diffusivity and sticking characteristics
(diffusion limited or reaction limited cluster aggregation, for
example) but, in principle, the aggregation eventually results in a
phase separation. The formation of metastable finite-size clusters
is not contemplated by the theory, but is the result of some
different mechanisms. Usually, a metastable cluster phase is
considered the consequence of a 'frustrated aggregation' due to the
decreasing diffusivity that progressively reduces the chance of an
encounter for the larger clusters. Then the metastable cluster phase
is 'kinetically stabilized'. However, within this theoretical
framework, it is difficult to account for the dependence of the
aggregate size on their net residual charge.

A different mechanism of stabilization of a cluster phase has
been described in the case of colloids characterized by the
presence of short-range attractive and long-range repulsive
interactions
\cite{Groenewold01,Sciortino04b,Bordi05b,Bartlett05,
Bartlett05b}. Different authors have hypothesized that stable
clusters can result from the equilibrium of long-ranged screened
electrostatic repulsions and short-range attractions, where the
like-charged primary particles within the same clusters repel
each other even though they are far apart and not in a direct
contact. However, due to the very small screening length in
aqueous systems, when compared to the size of lipid vesicles,
these model does not seem applicable to polyelectrolyte-liposome
suspensions.

We have recently shown that the formation of a cluster phase and
also the observed increase, with the aggregate size, of the
apparent cluster-cluster repulsive interactions as measured by
the second virial coefficient \cite{Bordi07c}, can be
effectively taken into account  the aggregation of the
polyelectrolyte-decorated particles as a thermally activated
process.

Velegol and Twar \cite{Velegol01} have recently developed an
analytical model for the potential of mean force between
non-uniformly charged particles. They have shown that a non-uniform
charge distribution at the surface of colloidal particles gives
place to an inter-particle potential that, even in the case of
same-sign charged particles, has an attractive component. The model
is based on the Derjaguin approximation and on an extension of the
Hogg-Healy-Fuerstenau [HHF] model \cite{Hogg66}.

The resulting potential depends on the values of the average
electrostatic surface potential, $\Phi$,   and on its standard
deviation, $\sigma$. This inter-particle potential, for some
combinations of the values of $\Phi$ and $\sigma$, shows a
maximum, close to the particles surface, representing an energy
barrier that particles have to pass for sticking together. The
height $H$ of this barrier increases with the radius of
curvature, $R$, of the surface of the two approaching particles,
i.e., for deformable particles (as is the case of
polyelectrolyte decorated lipid vesicles), with the aggregate
size, and depends on the surface potential $\Phi$, going
approximately as $H \propto R^2 $. By adding to this potential a
contribution from van der Waals attraction, operating at smaller
distance, this picture does not result substantially modified.

As a further test of this picture, in this paper, we investigate
the effect of the temperature on the aggregation behavior of
colloidal charged particles induced by addition of oppositely
charged polyions. We will show that, as is expected for a
thermally activated process, the size of the aggregates
increases when they are formed at higher temperatures. Here, we
deal with the complexation between an anionic linear polyion,
sodium polyacrylate [NaPA], and liposomes built up with the
cationic lipid di-oleoyl-trimethyl-ammonium-propane [DOTAP]
carried on at different temperatures, from 5 to 80 $^{\circ}$C.
Although the range of temperature variation is intrinsically
limited by the liposome stability and by the aqueous nature of
the dispersing medium, a significant effect is observed on
cluster properties, particularly on their size that, close to
the isoelectric point, increases appreciably when the
temperature is increased from 5 to 80 $^{\circ}$C.

\section{Experimental}
\subsection{Materials}
Positively charged liposomes were prepared by employing the
cationic lipid di-oleoyl-trimethyl-ammonium-propane [DOTAP],
purchased from Avanti Polar Lipids (Alabaster, AL) and used
without further purification. The negatively charged
polyelectrolyte sodium poly\-acrylate,  $[-CH_2CH(CO_2Na)-]_n$
[NaPAA], with nominal molecular weight 60 kD,   was purchased
from Polysciences Inc. (Warrington, PA) as 25 $\%$ aqueous
solution. All liposomal samples and polyelectrolyte solutions
were prepared in Milli-Q water, with electrical conductivity
less than 1$\cdot$10$^{-6}$ mho/cm.

\subsection{Preparation of cationic lipid-polyion complexes}
Liposomes were prepared by dissolution of an appropriate amount
of DOTAP in  methanol-chloroform solution (1:1 vol/vol). After
overnight vacuum roto-evaporation of the solvent, the dried
lipid film was re-hydrated with Milli-Q quality water. The
re-hydration process was carried out for 1 hour at a temperature
of $40^{\circ}$C, well above the main phase transition
temperature of this lipid ($T_f \approx 0 ^{\circ}$C). In order
to form small uni-lamellar vesicles, the lipid solution was
sonicated for 1 hour at a pulsed power mode, until the solution
appeared optically transparent in white light; the solution was
then filtered by means of a Millipore 0.4 $\mu m$ polycarbonate
filter. For all the experiments, liposomes were prepared at a
concentration of 1.5 mg/ml, corresponding  to approximately
1.5$\cdot$10$^{13}$ particle/ml  (average radius 40 nm) and the
solutions were stored at $4^{\circ}$C. For all the preparations,
the size distribution of the DOTAP liposomes was log-normal with
a mean hydrodynamic radius $40\pm 5 $ nm and with a
poly-dispersity of the order of 0.2, as expected for a rather
homogeneous particle suspension.

Measurements on liposomes-polyelectrolyte complexes were performed
at various temperatures, from 5 to 80 $^{\circ}$C. Lipoplexes were
formed immediately before each measurement by mixing equal volumes
of the liposome suspension and polyion solutions at the appropriate
concentration, as elsewhere described \cite{Bordi06,Sennato05BBA}.
Before mixing, liposomes and polyelectrolyte solutions were
thermostatted at the desired temperature. The mixed sample was
placed in a thermostatted cell for the measurement of both
electrophoretic mobility and size and size distribution.
Electrophoretic mobility measurements were performed 5 minutes after
the mixing, immediately followed by size determination. The
temperature was controlled within $\pm 0.1^{\circ}$C.

\subsection{Electrophoretic mobility and Dynamic light scattering measurements}
Both the electrophoretic mobility and the size and size
distribution of the suspended particles were measured by means
of an integrated apparatus, NanoZetaSizer (Malvern Instruments
LTD, UK) equipped with a 5 $mW$ He-Ne laser. In this way the
electrophoretic parameters and the size of the particles can be
measured on the same sample and almost simultaneously, reducing
the experimental uncertainties related to sample preparation,
thermal gradients and convective movement that, in the presence
of large aggregates within the suspension, can be considerable.

The electrophoretic mobility measurements were carried out by means
of the laser Doppler electrophoresis technique. The mobility $u$ was
converted into the $\zeta$-potential using the Smoluchowski relation
$\zeta=u \eta/\epsilon$, where $\eta$ and $\epsilon$ are the
viscosity and the permittivity of the solvent phase, respectively.

Size and size distribution of liposomes and polyions-liposome
aggregates were measured by means of dynamic light scattering (DLS)
technique, collecting the normalized intensity autocorrelation
functions at an angle of $173^{\circ}$ and analyzing the collected
data by using the CONTIN algorithm \cite{Provencher82}, in order to
obtain the decay time distribution of the electric field
autocorrelation functions.  Decay times are used to determine the
distribution of the diffusion coefficients $D$ of the particles,
which in turn can be converted in a distribution of apparent
hydrodynamic radii $R_H$ using the Stokes–Einstein relationship $R_H
= K_B T/6 \pi \eta D$, where $K_BT$ is the thermal energy and $\eta$ the solvent viscosity. The
values of the radii shown here correspond to the average values on
several measurements and are obtained from intensity weighted
distributions.

\subsection{Simulations}
We simulated the aggregation behavior of a system composed by $N_p$
= 10000  particles of initial diameter $2R = 80\;nm$ in a cubic box
of volume $V$ with packing fraction $\phi = 4\pi \rho R^3/3 = 0.01$,
where $\rho = N_p/V$ is the number density. We carried out MC
simulation using local metropolis algorithm at T = 298 K. Particles
interact via a short-range potential defined by eq.
\ref{eq:potVelegol}. The parameters of the simulations have been
detailed elsewhere \cite{Truzzolillo08}. We only recall here that to
incorporate a Brownian dynamics in the MC algorithm, the
\textit{i}th particle is selected with a probability proportional to
$R_0/R_i$, where $R_0$ is the initial radius and $R_i$ is the radius
of the \textit{i}th aggregate. Each selected aggregate is then moved
in each direction by a random quantity (uniformly distributed
between ±0.2 nm). Since the aggregation process progressively slows
down, simulations were interrupted when a plateau in the time
dependence of the aggregate average radius was reached.

\section{Results and discussion}
We have investigated by means of electrophoretic and dynamic light
scattering measurements the complexation of cationic DOTAP liposomes
in the presence of anionic NaPA polyelectrolytes, as a function of
the temperature in the interval from 5 to $80\,^{\circ}$C.

Preliminarily, the thermal behavior of the liposome dispersion in
the absence of the polyelectrolyte was investigated. The temperature
of the liposome dispersion was varied by one  step, by immersing the
samples, at the initial temperature of T=25 $^{\circ}$C, in a
thermostatted bath at the desired final temperature and allowing to
thermalize for 30 minutes before measuring the size. Following this
procedure, no significant variation of liposome size is observed due
to temperature change. It was found that different thermal protocols
induce large variations of liposome size, as it was observed for
other liposome systems \cite{Cinelli07}. For example, by varying the
temperature very slowly, in steps of $1\div2\,\,^{\circ}$C, and
allowing to thermalize for several minutes after each step, we have
observed a decrease in the vesicles radius as large as 50\% in
passing from T=5 to T=80 $^{\circ}$C (data not shown).  A small, but
significant, increase of the $\zeta$-potential was observed at
increasing temperature, independently of the procedure employed to
vary the temperature (data not shown). Taking the advantage of these
preliminary investigations, only the the procedure of 'rapid
immersion' was employed.

The stability of DOTAP liposome maintained at the desired
temperature within the interval 5$\div$80 $^{\circ}$C was checked by
repeated measurements of size and $\zeta$-potential. Over a period
of 24 hours nor the size nor the $\zeta$-potential show appreciable
variations. This high stability of DOTAP liposomes was largely
expected, since in the whole  interval of temperatures investigated
the double layers of this lipid are in the liquid phase (the main
transition temperature of this lipid is close to 0 $^{\circ}$C) and
do not undergo any phase transition that could facilitate a lipid
restructuring \cite{Hirsch99}.

Analogously, also the complexes that form on mixing  liposomes
with the polyelectrolyte remain stable in time over more than 12
hours, exhibiting a mono-modal log-normal distribution when
analyzed by CONTIN. Only very close to the isoelectric point,
large clusters appear, that rapidly precipitate, this phenomenon
being more pronounced at the higher temperatures investigated.

As usual \cite{Sennato05BBA,Acosta08}, the behavior of the
$\zeta$-potential and hydrodynamic radius of complexes is shown as a
function of a 'stoichiometric' charge ratio $\xi$, defined as the
ratio between the total number $N^-$ of the negative charges on the
polyion chains and the total number $N^+$ of the positive charges on
DOTAP molecules in the whole suspension
\begin{equation}\label{eq:ChargeRatio}
    \xi=\frac{N^-}{N^+}=\frac{C_M}{Mw_{M}}\frac{Mw_{D}}{C_D}
\end{equation}
where $C_M$ and $C_D$ are the (weight) concentrations of the polyion
and DOTAP, respectively, and $Mw_{M}$ and $Mw_{D}$ are the molecular
weights of the repeating unit of the polyion and of DOTAP.  This
definition of $N^- / N^+$ considers all the charges on the
liposomes, both on the outer and on the inner leaflet of the
vesicle.

Fig. \ref{fig:1} shows the hydrodynamic diameter $2R$ and the
corresponding $\zeta$-potential as a function of $\xi$ for the
liposome-polyelectrolyte complexes, at some selected temperatures in
the range from 5 to 80 $^{\circ}$C. At all the temperatures
investigated, liposome-polyelectrolyte complexes show a 'reentrant
condensation' accompanied by a $\zeta$-potential inversion, i.e. the
inversion of the sign of their net electric charge. Close to the
inversion point (isoelectric point), the complexes reach their
maximum size. However, at any given polyelectrolyte-liposome ratio,
the size of the aggregates depends on the temperature, the largest
clusters being observed at the higher temperatures. This result
indicates that aggregation is favored by higher temperatures, as it
should be in a thermally activated process.

We have recently shown that the observed reentrant condensation
and the accompanying charge inversion can be both explained in
terms of the correlated adsorption of polyelectrolytes on the
oppositely charged liposomes \cite{Sennato05BBA,Bordi06}. The
resulting attraction between the polyelectrolyte-decorated
particles is due to the non-uniform distribution of the net
electrostatic charge at their surface \cite{Bordi07c,
Truzzolillo08}.

The adsorption of the linear polyelectrolytes on the surface of the
oppositely charged particles occurs in a highly correlated manner.
In fact, the competing interactions between polymer and surface
(attraction) and between polymers (repulsion) result in a locally
ordered, correlated adsorption of the polymer
\cite{Mou95,Nguyen00,Dobrynin00,Grosberg02,Pericet04}.

The non-uniform distribution of the
adsorbed chains gives also rise to a non-uniform distribution of the
surface electric charge, with alternate patches where the charge of
the polymer or the one  of the particle surface are locally in
excess.

Velegol and Twar \cite{Velegol01} have recently developed an
analytical model for the potential of mean force between
non-uniformly charged particles. They have shown that a non-uniform
charge distribution at the particle surface results in an
inter-particle potential that, even in the case of particles that
bear a net charge of the same sign, has an attractive component. The
model is based on an extension of the Hogg-Healy-Fuerstenau [HHF]
model \cite{Hogg66} and on the Derjaguin approximation. This
approximation  holds when the  characteristic radius of the smaller
sphere is much larger than the characteristic length scale for the
interaction \cite{Todd04}, i.e. the Debye screening length.


By using the Derjaguin approximation, the generic force $F(h)$
between the surfaces of two spheres of radii $R_1$ and $R_2$, at a
distance $h$  in terms of the potential $G(h)$ that would be
observed  the two surfaces were infinite planes at the same distance
$h$, can be written as \cite{Todd04}
\begin{equation}\label{eq:derjaguin}
    F(h) \propto \frac{R_1 R_2}{R_1+R_2} G(h)
\end{equation}
This expression clearly shows that, as a general rule and
independently of the nature of the inter-particle potential,
whenever the Derjaguin approximation holds, when their radius
increases, the force between two spherical particles also increases,
toward the limiting force observed for two planes facing each other.

According to Velegol and Twar,  the mean force pair interaction
potential between two spherical particles (A and B) bearing a
non-uniformly distributed electric charge on their surface can be
written, in units of the thermal energy  $k_B T$, as
\begin{multline}\label{eq:potVelegol}
    \langle \Phi \rangle=\frac{\epsilon \pi R_A R_B}{R_A+R_B}\\
    \left [(\zeta_A^2+\zeta_B^2 +\sigma_A^2+\sigma_B^2)
    \ln(1-e^{-2\kappa H})+2\zeta_A \zeta_B \ln \left(\coth\frac{\kappa H}{2}\right) \right]
\end{multline}
were $H$ is the distance between the surfaces of the two approaching
particles, $\epsilon$ the permittivity of the medium and
$\kappa^{-1}$ the Debye screening length. This potential depends on
the values of the average electrostatic surface potentials,
$\zeta_i$-potential, and on their standard deviations $\sigma_i$,
combining a net charge-dependent monopole term (for $\zeta_i\neq
0$), which is repulsive for like-charged particles, and an always
attractive multipole term ($\sigma_i\neq 0$) arising from the charge
heterogeneity. These two components lead up to a global maximum and
to a potential barrier. The height of the potential barrier that two
approaching particles must overcome in order to stick together and
the separation $H_{max}$ between the particle surface at which this
maximum occurs can be evaluated from eq. \ref{eq:potVelegol}. For
two identical particles ($R_A = R_B = R$) we obtain
\begin{equation}
\label{eq:PhiMax}
\Phi_{max}= \pi \epsilon R \left\{ (\zeta^2 + \sigma^2) \ln \left[1-
\left( \frac{\zeta^2}{\zeta^2+\sigma^2} \right)^2 \right]+ \zeta^2
\ln \left[\frac{2 \zeta^2+\sigma^2}{\sigma^2}  \right] \right\}
\end{equation}
 and
\begin{equation}\label{eq:Hmax}
    H_{max}= \frac{1}{\kappa}\ln \left(  \frac{\zeta^2+\sigma^2}{\zeta^2} \right)
\end{equation}
respectively. The height of this barrier  increases with the radius
of curvature, $R$, of the surface of the two approaching particles,
i.e., for deformable particles (as is the case of polyelectrolyte
decorated lipid vesicles \cite{Bordi07c}) with the aggregate size.

At each temperature, the increase of the height of the potential
barrier favors the formation of finite size aggregates.
 Within this framework,
the observed increase of the size of the aggregates at the
higher temperatures  can be easily explained. In fact, as the
temperature is increased, the fraction  of particles that have a
sufficient thermal energy to overcome the potential barrier
increases. Since the barrier height itself increases with the
size of the aggregates, larger and larger aggregates are
stabilized at higher temperatures.

The aggregation behavior of a system of non-uniformly charged
particles was recently investigated by the present authors by
means of Monte Carlo simulations \cite{Truzzolillo08} in order
to derive the limiting size of the aggregates. Simulations
qualitatively reproduce the observed trends in different
colloidal systems. As an example, in fig. \ref{fig:D1}, it is
shown the typical time evolution of the normalized mean radius
of the clusters that are formed by non-uniformly charged
particles interacting trough the potential described by equation
\ref{eq:potVelegol}. Fig. \ref{fig:D1} shows the effect on this
evolution, for two different average surface electrostatic
potential $\zeta$= 15 mV and $\zeta$=11 mV, for a fixed value of
the non-uniformity parameter $\sigma$= 15 mV. After an initial
transient regime the radius of the aggregates increases in a
similar way for the two considered $\zeta$-potential values. The
growth process slows down at long times and the aggregate size
reaches a long-time limit value that depends on the electrical
surface parameters $\zeta$ and $\sigma$. For the same value of
surface inhomogeneity $\sigma$, the kinetic arrest is reached at
higher values of mean cluster size when surface potential
$\zeta$ is lower (see inset of Fig. \ref{fig:D1}).

In all our simulations the plateau is reached when the height of
the potential barrier between the particles is $\approx 10 K_B
T$. This observation provides a strong support to the hypothesis
that the dynamics slows down and the arrest of the growth
observed on the timescale sampled by our simulation is
controlled by a mechanism of 'thermal stabilization', i.e., the
increase of the potential barrier with the radius  stabilizes a
'limiting size' of the aggregates. Fig. \ref{fig:D2} shows the
comparison between the mean cluster size reached at the kinetic
arrest obtained by MC simulations (filled circles) and the ones
calculated according to eq. \ref{eq:PhiMax} within the framework
of Velegol and Twar model, with $\Phi_{max}$= 10 $K_B$T, as a
function of $\zeta$ and as a function of $\sigma$ (inset, Fig.
\ref{fig:D2}). The observed agreement supports the evidence for
a kinetic arrest at a well defined value of repulsive
interactions between particles.

Here, we only report on the effect of the temperature on the
size of the aggregates. Fig. \ref{fig:D} shows the limiting
radius  of the aggregates, as a function of $\zeta$-potential,
according to the eq. \ref{eq:PhiMax}, at the different
temperatures. In the inset, the effect of the charge
inhomogeneity parameter $\sigma$ on the mean force potential is
shown as a function of the minimal distance $H$ between the
surfaces of the two approaching particles. These behaviors
justifies the increase of the size of the clusters as the
temperature is raised for all the values of the surface
potential.

The experimental results shown in Fig. \ref{fig:1} can be
compared in a semi-quantitative way with the predictions of the
model. Expressing the  $\zeta$-potential  as a function of the
polyelectrolyte-liposome charge ratio $\xi$ (Fig. \ref{fig:1},
bottom panel), the predicted average radius of the aggregates
can be calculated as a function of $\xi$ from eq.
\ref{eq:PhiMax}.

Charge inhomogeneity is more difficult to evaluate. Because of
the irregular shape and of the high deformability of our
aggregates, a direct measurement of the $\sigma$ values  by
electrophoretic rotation \cite{Velegol02} appears difficult.
However, a dependence of $\sigma$ on $\xi$ can be hypothesizes
on the basis of simple considerations. $\sigma$ is expected to
be close to zero at the beginning of the aggregation process,
where the aggregates show a $\zeta$-potential very similar to
the one of bare primary particles. At these values of the charge
ratio, only a few polyion chains adsorb individually on the
liposome surface and these isolated spots produce negligible
inhomogeneity on the average. The inhomogeneity of the
adsorption layer increases with the increase of the polyion
content, and $\sigma$ is expected to reach its maximum value at
the isoelectric point, and  then it is expected to decrease
again as the adsorbed polyion layer becomes more and more
compact. This behavior can be modeled assuming a Gaussian
dependence of $\sigma$ on $\xi$, whose width is calculated as
the half-height width of the condensation region at the
different temperatures (Fig. \ref{fig:1} upper panels), while
the amplitude is a free parameter to be determined by the
fitting procedure. Finally, the value of the potential barrier
height $\Phi_{max}$ in eq. \ref{eq:PhiMax} is somewhat
arbitrarily fixed at 10 $K_B$T. This choice appear justified by
MC simulations, showing that for this height a dynamical arrest
of the aggregation occurs (see, for example, Fig. \ref{fig:D1},
\ref{fig:D2} and ref. \cite{Truzzolillo08}). With all these
assumptions, the eq. \ref{eq:PhiMax} can be inverted and the
expected limiting radius calculated in the whole range of the
charge ratio $\xi$ investigated.

Fig. \ref{fig:2} shows a typical result of this analysis. The
 diameters measured at 20 $^\circ$C (symbols) are
compared with the $R(\zeta, \sigma)$ curve calculated on the basis
of the eq. \ref{eq:PhiMax} (continuous line). An analogous
reasonable agreement is found for all the  temperatures investigated
(data not shown).

Interestingly, the amplitude of the gaussian curve that
represents the dependence of $\sigma$ on $\xi$, i.e. the maximum
value assumed by $\sigma$, does not vary significantly with the
temperature (the values being comprised within the interval $50
\pm 5 $ mV), indicating that inhomogeneity of the surface
polyion coverage at the isoelectric point is largely independent
of temperature.

It must be noted that the final cluster size has an explicit
dependence on temperature through the expression of the
potential barrier height (eq. \ref{eq:PhiMax}). It could be
interesting to determine if temperature changes affect the
cluster size according to what is predicted by the explicit
dependence on the temperature $T$, or if other effects have to
be considered. To this aim, we have plotted together all the
data at the different temperatures in Fig. \ref{fig:3}, shifting
the position of the aggregation peaks respect to the one at 80
$^{\circ}$C. Fig. \ref{fig:3} shows that all the experimental
data, in the whole range 5$\div$80 $^\circ$C, collapse on a
master curve when re-scaled, suggesting that the aggregation
kinetics of polyion-decorated liposome in the temperature range
investigated is the same, independently of temperature, and it
is typical of a thermal activated process.

This result is expected within the framework of Velegol and Twar
model \cite{Velegol02}. Considering the dependence of $R$ on $\zeta$-potential and $\sigma$,
(eq. \ref{eq:PhiMax}), once that the dependence $\zeta(\xi)$ and
$\sigma(\xi)$ are evaluated at each given temperature with the aid of experimental data,
only the linear dependence on temperature, in
the potential barrier height, remains.

At the isoelectric point, where large flocs are observed, higher
temperatures seem to promote a more rapid gross aggregation. The universality of the
aggregation appears to be valid within the condensation region but not at the
isoelectric point. In fact, data of clusters with almost zero
$\zeta$-potential  fail to collapse on the master curve,  due to
large fluctuations of the measured size at the isoelectric point, where $\zeta$-potential has
vanishing values.

Notably, at the lower temperature investigated, 5 $^{\circ}$C,
on the same experimental time scale employed in the
other measurements, flocculation at the isoelectric point is not
observed and almost-neutral clusters remain stable.

The charge ratio $\xi$ where the maximum size of the aggregates
is observed does not vary significantly at the different
temperatures (upper panels, fig. \ref{fig:1}), as well as the
position of $\zeta$-potential sign reversal, corresponding to
charge inversion of clusters (bottom panels, Fig. \ref{fig:1}).

The increase of temperatures (bottom panels, Fig. \ref{fig:1})
promotes a slight increase of the $\zeta$-potential of the
complexes at charge ratios below the isoelectric point. This
effect is better observed in Fig. \ref{fig:ZT}, where the data
are properly shifted to make the different positions of the
isoelectric point coincident.
In the regions below and above  the isoelectric
point, $\zeta$-potential increases with temperature. This small
increase can be explained as the result of two combined effects.

First, the effect of counterion condensation on both
polyelectrolytes and liposomes must be taken into account.
Different theoretical models hypothesize that the counterions,
on approaching to oppositely charged surface, 'condensed' around
a highly charged linear polyelectrolyte would not necessarily be
released because of entropic effects
\cite{Sens00,Dobrynin00,Nguyen01b}. Recently, these hypotheses
have also been substantiated by experimental evidence
\cite{Bordi07b,Radeva06,Radeva07}. At the higher temperature,
the increasing entropy of counterions reduces the effects of
'condensation' \cite{Belloni98} so that the effective charge of
the bare surface increases.
The second effect is connected to the conformational entropy of
the adsorbed polyions, that with the increase of the polyion
content, and hence of the surface coverage, becomes relevant. In
fact, at low temperatures, polyions tends to assume a more flat
conformation on liposome surface, thus contributing a stronger
screening effect, which reflects in the lower values of the
$\varsigma$-potential observed. At higher temperatures, polyions
assume a more disordered conformation with more loops. A similar
mechanism can be invoked to interpret the behavior observed
above the isoelectric point. At charge ratios close to 1.5 above
the isoelectric condition, a sensible increase of the
$\zeta$-potential (in absolute values) with the temperature is
observed. At T=80 $^{\circ}$C above $\xi\approx 1.5$,
$\zeta$-potential values begin to stabilize, as if a maximum
polyion adsorption were reached. Again, this behavior could be
connected with the temperature-depending adsorption properties
of polyions, when, with the  increase of the  thermal energy,
polyion chains are less strongly bound to the oppositely charged
surface, until a desorption limit is reached
\cite{Muthukumar87}.

\section{Conclusion}

We have investigated the effect of temperature on the reentrant
condensation of polyelectrolyte-liposome complexes, in a temperature
range from 5 to 80 $^\circ$C.
At all the temperatures investigated, the observed phenomenology is similar, with
the formation of polyion-liposome complexes which aggregate in large
clusters close to the isoelectric point. Temperature does not modify
the overall aspect of the phenomena.
However,  significant changes of size and
$\zeta$-potential values of the polyion-decorated liposome aggregates are observed,
connected to the increase of temperature.

The observed effects can
be interpreted in terms of a thermally activated process,
where the potential barrier that controls the aggregation arises
from the combined effect of the electrostatic repulsions, due to the
residual net charge on the primary particles (the polyelectrolyte-decorated
liposomes), and of an attractive term due to the non-uniformity of the
adsorbed chains and hence of the surface electric potential.

The observed behavior is interpreted within the framework of Velegol
and Twar model \cite{Velegol01}, that takes into account the effect
of charge anisotropy on aggregating particles. A good agreement is
found between calculated and experimental values of cluster size, for
all the temperatures investigated. More interestingly, the analysis
according to the Velegol and Twar model evidences that, once the
dependence of radius on temperature is taken into account,
interactions of oppositely charged polyelectrolytes and liposomes
show universality properties. In fact, cluster sizes within the
reentrant condensation region collapse on a single master curve,
thus excluding the presence of non-electrostatic interactions or
effects connected with cluster growth rate on the observed
aggregation.



\newpage


\providecommand{\refin}[1]{\\ \textbf{Referenced in:} #1}
\begin{thebibliography}{10}

\bibitem{Raspaud99}
Raspaud,~E.;\ \ Chaperon,~I.;\ \ Leforestier,~A.;\ \ Livolant,~F.
  \textit{Biophys. J.} \textbf{1999,} \textsl{77,} 1547-1555.

\bibitem{Wang00}
Wang,~Y.;\ \ Kimura,~K.;\ \ Dubin,~P.~L.;\ \ Jaeger,~W. \textit{Macromolecules}
  \textbf{2000,} \textsl{33,} 3324-3331.

\bibitem{Bordi04}
Bordi,~F.;\ \ Cametti,~C.;\ \ Diociaiuti,~M.;\ \ Gaudino,~D.;\ \ Gili,~T.;\ \
  Sennato,~S. \textit{Langmuir} \textbf{2004,} \textsl{20,} 5214-5222.

\bibitem{Bordi06}
Bordi,~F.;\ \ Cametti,~C.;\ \ Sennato,~S.;\ \ Diociaiuti,~M. \textit{Biophys.
  J} \textbf{2006,} \textsl{91,} 1513-1520.

\bibitem{Yaroslavov07}
Yaroslavov,~A.;\ \ T.,~S.;\ \ Rakhnyanskaya,~A.;\ \ Ermakov,~Y.;\ \
  Burova,~T.;\ \ Grinberg,~V.~Y.;\ \ Menger,~F.~M. \textit{Langmuir}
  \textbf{2007,} \textsl{23,} 7539-44.

\bibitem{Yaroslavov07b}
Sybachin,~A.~V.;\ \ Efimova,~A.~A.;\ \ Litmanovich,~E.~A.;\ \ Menger,~F.~M.;\ \
  Yaroslavov,~A. \textit{Langmuir} \textbf{2007,} \textsl{23,} 10034-10039.

\bibitem{Radeva07}
Kamburova,~K.;\ \ Radeva,~T. \textit{J. COLLOID INTERF. SCI.} \textbf{2007,}
  \textsl{313,} 398-404.

\bibitem{Volodkin07}
Volodkin,~D.;\ \ Ball,~V.;\ \ Schaaf,~P.;\ \ Voegel,~J.-C.;\ \ Mohwald,~H.
  \textit{Biochim. Biophys. Acta} \textbf{2007,} \textsl{1768,} 280-290.

\bibitem{Macdonald07}
Pozharski,~E.~V.;\ \ R.C.,~M. \textit{Mol. Pharmaceutics} \textbf{2007,}
  \textsl{4,} 962-974.

\bibitem{Bordi07c}
Bordi,~F.;\ \ Cametti,~C.;\ \ Sennato,~S.;\ \ Truzzolillo,~D. \textit{Phys.
  Rev. E} \textbf{2007,} \textsl{76,} 061403-12.

\bibitem{Borkovec08}
Gillies,~G.;\ \ Lin,~W.;\ \ Borkovec,~M. \textit{J. Phys. Chem. B}
  \textbf{2007,} \textsl{111,} 8626-8633.

\bibitem{Sennato08}
Sennato,~S.;\ \ Bordi,~F.;\ \ Cametti,~C.;\ \ Marianecci,~C.;\ \ Carafa,~M.;\ \
  Cametti,~M. \textit{J. Phys. Chem. B} \textbf{2008,} \textsl{112,} 3720-27.

\bibitem{Mou95}
Mou,~J.;\ \ Czajkowsky,~D.~M.;\ \ Zhang,~Y.;\ \ Shao,~Z. \textit{FEBS Letters}
  \textbf{1995,} \textsl{371,} 279-282.

\bibitem{Nguyen00}
Nguyen,~T.~T.;\ \ Grosberg,~A.~Y.;\ \ Shklovskii,~B.~I. \textit{Phys. Rev.
  Lett.} \textbf{2000,} \textsl{85,} 1568-1571.

\bibitem{Dobrynin00}
Dobrynin,~A.~V.;\ \ Deshkovski,~A.;\ \ Rubinstein,~M. \textit{Phys. Rev. Lett.}
  \textbf{2000,} \textsl{84,} 3101-3104.

\bibitem{Grosberg02}
A.~Y.~Grosberg,~T. T.~Nguyen,~B. I.~S. \textit{Rev. Mod. Phys.} \textbf{2002,}
  \textsl{74,} 329-345.

\bibitem{Pericet04}
Pericet-Camara,~R.;\ \ Papastavrou,~G.;\ \ Borkovec,~M. \textit{Langmuir}
  \textbf{2004,} \textsl{20,} 3264-71.

\bibitem{LaMer63}
La~Mer,~V.~K.;\ \ Healy,~T.~W. \textit{Rev. Pure Appl. Chem.} \textbf{1963,}
  \textsl{13,} 112-133.

\bibitem{Mowald01}
Ahrens,~H.;\ \ Baltes,~H.;\ \ Schmitt,~J.;\ \ M\"{o}hwald,~H.;\ \ Helm,~C.~H.
  \textit{Macromolecules} \textbf{2001,} \textsl{34,} 4504-4512.

\bibitem{Henon03}
Vagharchakian,~L.;\ \ Hénon,~S. \textit{Langmuir} \textbf{2003,} \textsl{19,}
  7989-7994.

\bibitem{Dobrynin01}
Dobrynin,~A.~V.;\ \ Deshkovski,~A.;\ \ Rubinstein,~M. \textit{Macromolecules}
  \textbf{2001,} \textsl{34,} 3421-3436.

\bibitem{Li06}
Li,~Z.;\ \ Wu,~J. \textit{Phys. Rev. Lett.} \textbf{2006,} \textsl{96,} 048302.

\bibitem{Henon04}
Vagharchakian,~L.;\ \ Desbat,~B.;\ \ Hénon,~S. \textit{Macromolecules}
  \textbf{2004,} \textsl{37,} 8715-8720.

\bibitem{Leong95}
Leong,~Y.~K.;\ \ Scales,~P.~J.;\ \ Healy,~T.~W.;\ \ Boger,~D.~V.
  \textit{Colloids Surfaces A} \textbf{1995,} \textsl{95,} 43-52.

\bibitem{Gregory73}
Gregory,~J. \textit{J. Colloid Interface Sci.} \textbf{1973,} \textsl{42,}
  448-456.

\bibitem{Miklavic94}
Miklavic,~S.~J.;\ \ Chan,~D. Y.~C.;\ \ R.,~W.~L.;\ \ Healy,~T.~W. \textit{J.
  Phys. Chem.} \textbf{1994,} \textsl{98,} 9022-9032.

\bibitem{Khachatourian98}
Khachatourian,~A. V.~M.;\ \ Wistrom,~A.~O. \textit{J. Phys. Chem. B}
  \textbf{1998,} \textsl{102,} 2483-2493.

\bibitem{Velegol01}
Velegol,~D.;\ \ Thwar,~P. \textit{Langmuir} \textbf{2001,} \textsl{17,}
  7687-7693.

\bibitem{Podgornik95}
Podgornik,~R.;\ \ {\AA}kesson,~T.;\ \ J\"{o}nsson,~B. \textit{J. Chem. Phys.}
  \textbf{1995,} \textsl{102,} 9423-9434.

\bibitem{Keren02}
Keren,~K.;\ \ Soen,~Y.;\ \ Ben~Yoseph,~G.;\ \ Yechiel,~R.;\ \ Braun,~E.;\ \
  Sivan,~U.;\ \ Talmon,~Y. \textit{Phys. Rev. Lett.} \textbf{2002,}
  \textsl{89,} 88103-88106.

\bibitem{Kabanov00}
Kabanov,~V. A.and~Sergeyev,~V.~G.;\ \ Pyshkina,~O.~A.;\ \ Zinchenko,~A.~A.;\ \
  Zezin,~A.~B.;\ \ Joosten,~J. G.~H.;\ \ Brackman,~J.;\ \ Yoshikawa,~K.
  \textit{Macromolecules} \textbf{2000,} \textsl{33,} 9587-9593.

\bibitem{Radler98}
R\"{a}dler,~J. O.~Koltover,~I.;\ \ Jamieson,~A.;\ \
  Salditt,~T.and~Safinya,~C.~R. \textit{Langmuir} \textbf{1998,} \textsl{14,}
  4272-4283.

\bibitem{Groenewold01}
Groenewold,~J.;\ \ Kegel,~W.~K. \textit{J. Phys. Chem. B} \textbf{2001,}
  \textsl{105,} 11702-11709.

\bibitem{Sciortino04b}
Sciortino,~F.;\ \ Mossa,~S.;\ \ Zaccarelli,~E.;\ \ Tartaglia,~P. \textit{Phys.
  Rev. Lett.} \textbf{2004,} \textsl{93,} 055701.

\bibitem{Bordi05b}
Bordi,~F.;\ \ Cametti,~C.;\ \ Diociaiuti,~M.;\ \ Sennato,~S. \textit{Phys. Rev.
  E} \textbf{2005,} \textsl{71,} 050401-4(R).

\bibitem{Bartlett05}
Campbell,~A.~I.;\ \ Anderson,~V.~J.;\ \ van Duijneveldt,~J.~S.;\ \ Bartlett,~P.
  \textit{Phys. Rev. Lett.} \textbf{2005,} \textsl{94,} 208301.

\bibitem{Bartlett05b}
Sanchez,~R.;\ \ Bartlett,~P. \textit{J. Phys.: Condens. Matter} \textbf{2005,}
  \textsl{17,} S3551-S3556.

\bibitem{Hogg66}
Hogg,~R.;\ \ Healy,~T.~W.;\ \ Fuerstenau,~D.~W. \textit{Trans Faraday Soc.}
  \textbf{1966,} \textsl{62,} 1638-51.

\bibitem{Sennato05BBA}
Sennato,~S.;\ \ Bordi,~F.;\ \ Cametti,~C.;\ \ Diociaiuti,~M.;\ \ Malaspina,~P.
  \textit{Biochim. Biophys. Acta} \textbf{2005,} \textsl{1714,} 11-24.

\bibitem{Provencher82}
Provencher,~S. \textit{Compt. Phys. Commun.} \textbf{1982,} \textsl{27,}
  213-227.

\bibitem{Truzzolillo08}
Truzzolillo,~D.;\ \ Bordi,~F.;\ \ Sciortino,~F.;\ \ Cametti,~C. \textit{arXiv
  cond-matt:0804.0781} \textbf{2008,} .

\bibitem{Cinelli07}
Cinelli,~S.;\ \ Onori,~G.;\ \ Zuzzi,~S.;\ \ Bordi,~F.;\ \ Cametti,~C.;\ \
  Sennato,~S.;\ \ Diociaiuti,~M. \textit{J. Phys. Chem. B} \textbf{2007,}
  \textsl{111,} 10032-39.

\bibitem{Hirsch99}
Hirsch-Lerner,~D.;\ \ Barenholz,~Y. \textit{Biochim. et Biophys. Acta}
  \textbf{1999,} \textsl{1461,} 47-57.

\bibitem{Acosta08}
Kang,~N.;\ \ Policova,~Z.;\ \ Bankian,~G.;\ \ Hair,~M.~L.;\ \ Zuo,~Y.~Y.;\ \
  Neumann,~A.~W.;\ \ Acosta,~E.~J. \textit{Biochim. Biophys. Acta}
  \textbf{2008,} \textsl{1778,} 291-302.

\bibitem{Todd04}
Todd,~B.;\ \ Eppell,~S.~J. \textit{Langmuir} \textbf{2004,} \textsl{20,}
  4892-4897.

\bibitem{Velegol02}
Feick,~J.~D.;\ \ Velegol,~D. \textit{Langmuir} \textbf{2002,} \textsl{18,}
  3454-3458.

\bibitem{Sens00}
Sens,~P.;\ \ Joanny,~J.-F. \textit{Phys. Rev. Lett.} \textbf{2000,}
  \textsl{84,} 4862-4865.

\bibitem{Nguyen01b}
Nguyen,~T.~T.;\ \ Shklovskii,~B.~I. \textit{J. Chem. Phys.} \textbf{2001,}
  \textsl{115,} 7298-7308.

\bibitem{Bordi07b}
Bordi,~F.;\ \ Cametti,~C.;\ \ Sennato,~S.;\ \ Viscomi,~D. \textit{Phys. Rev. E}
  \textbf{2006,} \textsl{74,} 030402-4(R).

\bibitem{Radeva06}
Milkova,~V.;\ \ Radeva,~T. \textit{J. Colloid Interf. Sci.} \textbf{2006,}
  \textsl{298,} 550-555.

\bibitem{Belloni98}
Belloni,~L. \textit{Coll. Surf. A} \textbf{1998,} \textsl{140,} 227-43.

\bibitem{Muthukumar87}
Muthukumar,~M. \textit{J. Chem. Phys.} \textbf{1987,} \textsl{86,} 7230-5.

\end{thebibliography}
\vspace{0.5cm}

\begin{figure}
\begin{center}
\includegraphics[width=16cm]{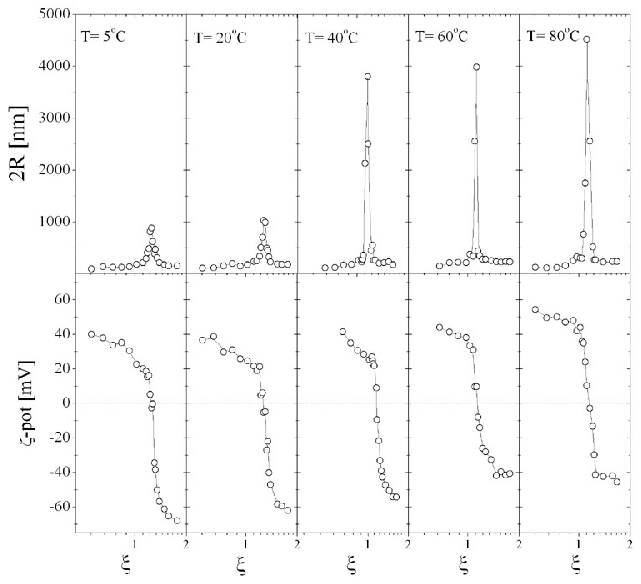}
\vspace{1cm} \caption{\label{fig:1}}
\end{center}
\end{figure}

Fig. \ref{fig:1} -  Reentrant condensation (aggregate diameter 2R,
upper panels) and charge inversion ($\zeta$-potential, bottom panels
) of cationic DOTAP liposomes in the presence of anionic NaPAA
polyelectrolyte. Data are shown as a function of the polyion-lipid
charge ratio parameter $\xi$ for the different temperatures
investigated. Lines connecting experimental data point guide eyes
only. It has to be noted that, at the isoelectrical point, due to
the intrinsic instability of the system, the data represent the
initial size or $\zeta$-potential values of flocculating aggregates.

\vspace{0.5cm}

\begin{figure}
\begin{center}
\includegraphics[width=13cm]{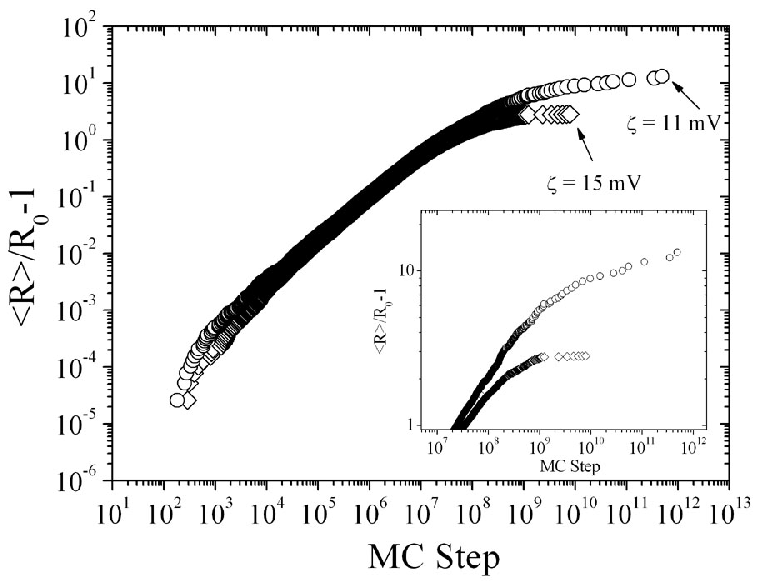}
\vspace{1cm} \caption{\label{fig:D1}}
\end{center}
\end{figure}

Fig. \ref{fig:D1} -  Some typical  MC-step evolution of normalized mean cluster radius
$<R>/R_0$ -1. Simulations have been carried out for different values of $\xi$ with a constant
value of the standard deviation $\sigma$=15 mV, $\diamond$: $\zeta$=15 mV, $\circ$: $\zeta$=11 mV.
The inset shows a magnification
of the evolution at longer times of the mean size , in the region where kinetical
 arrest is reached and mean cluster size is stable.

\vspace{0.5cm}

\begin{figure}
\begin{center}
\includegraphics[width=13cm]{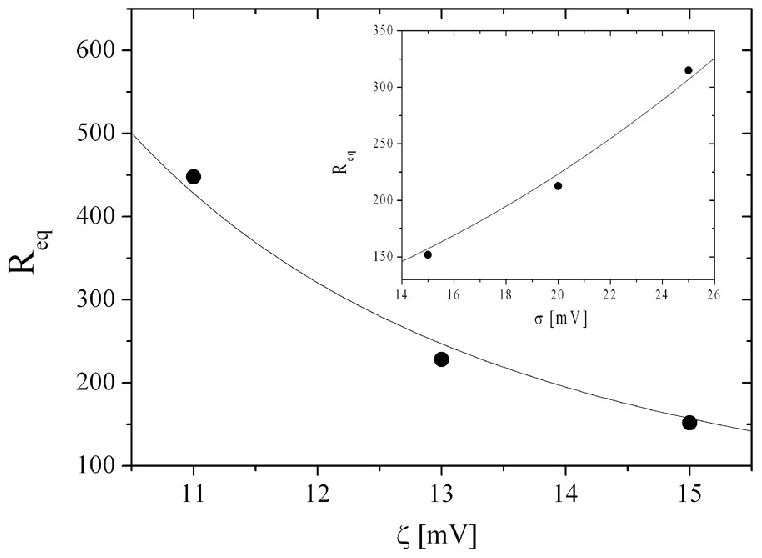}
\vspace{1cm} \caption{\label{fig:D2}}
\end{center}
\end{figure}

Fig. \ref{fig:D2} Comparison between mean cluster size at the kinetic arrest calculated according to the eq.
\ref{eq:PhiMax}
(continuous line) with $\Phi_{max}$=10 $K_BT$ and obtained by MC simulations (filled circles), as a function
of the electrostatic parameter $\zeta$ with fixed $\sigma$=15 mV. Inset shows mean cluster size according to eq.
$\ref{eq:PhiMax}$ (continuous line) and MC simulations (filled circles), at fixed $\zeta$=15 mV, as a function
of $\sigma$.
Each value of cluster size determined by MC simulations
is calculated as the mean value of the last five points obtained in
the MC simulations.

\vspace{0.5cm}

\begin{figure}
\begin{center}
\includegraphics[width=13cm]{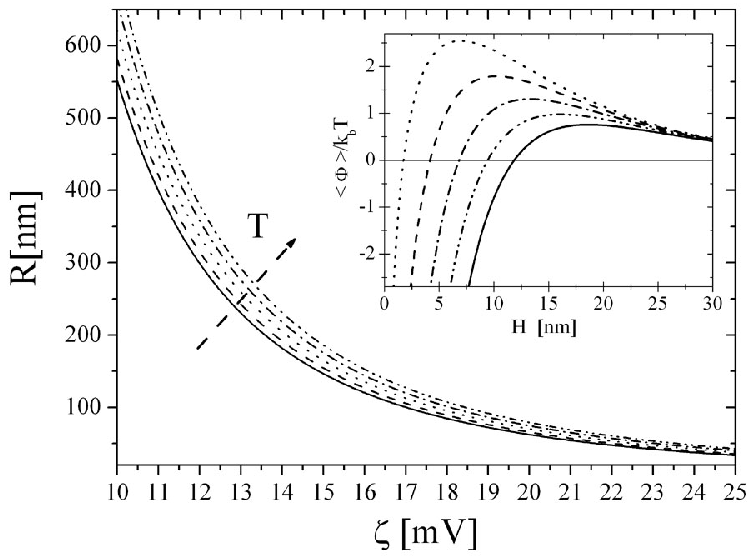}
\vspace{1cm} \caption{\label{fig:D}}
\end{center}
\end{figure}

Fig. \ref{fig:D} - Mean equilibrium radius of clusters at the
different investigated temperatures as a function of the
$\zeta$-potential, obtained by means of MC simulations according to
 eq. \ref{eq:PhiMax}, with $\Phi_{max}$= 10 $K_BT$ and $\sigma$= 15
mV (solid line: 5 $^{\circ}$C, dashed line: 20 $^{\circ}$C, dotted
line: 40 $^{\circ}$C, dash-dot line: 60 $^\circ$C, dash-dot-dot
line: 80 $^{\circ}$C). Inset shows potential curves as a function of
the minimal  distance between particle surfaces \textit{H}, at fixed
$\zeta$-potential (15 mV),  calculated for increasing  values of
$\sigma$ (solid line: 35 mV, dashed line: 30 mV, dotted line: 25
mV, dash-dot line: 20 mV, dash-dot-dot line: 15 mV).

\vspace{0.5cm}

\begin{figure}
\begin{center}
\includegraphics[width=13cm]{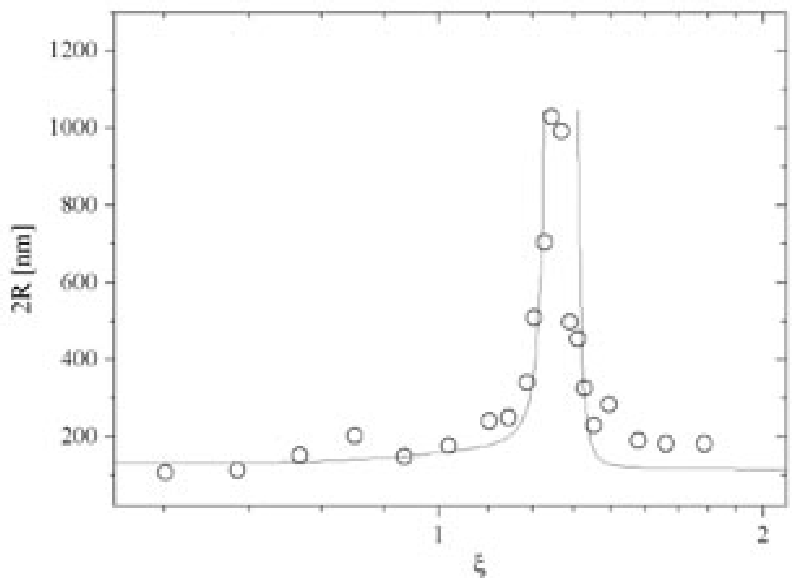}
\vspace{1cm} \caption{\label{fig:2}}
\end{center}
\end{figure}

Fig. \ref{fig:2} - Average hydrodynamic diameter of lipoplex
clusters at 20 $^{\circ}$C as a function of the polyion-lipid charge
ratio $\xi$. Line is calculated on the basis of eq. \ref{eq:PhiMax}
 within the Velegol and Twar model for inhomogeneously charged interacting colloidal particles.

\vspace{0.5cm}

\begin{figure}
\begin{center}
\includegraphics[width=13cm]{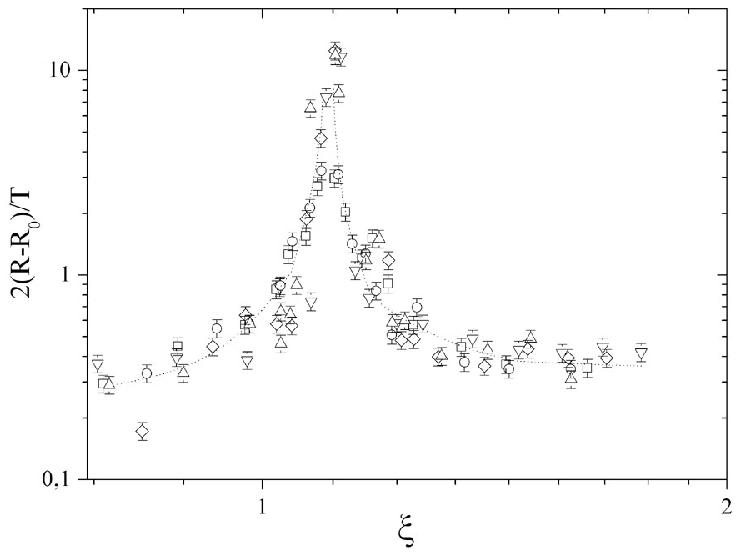}
\vspace{1cm} \caption{\label{fig:3}}
\end{center}
\end{figure}

Fig. \ref{fig:3} - Average hydrodynamic diameter of lipoplex
clusters normalized with respect to the temperature at which
measurement were performed,  for the various temperatures
investigated ($\square$:  5 $^{\circ}$C, ${\circ}$:  20 $^{\circ}$C,
$\bigtriangleup$:  40 $^{\circ}$C,
 $\bigtriangledown$: 60 $^{\circ}$C,  $\diamond$: 80 $^{\circ}$C).
Cluster size is rescaled with respect to the liposome size and
peak position is shifted with respect to the position at 80 $^{\circ}$C. Errors have been evaluated of the order
of 10 $\%$ of data. Line is  to guide eyes only.

\vspace{0.5cm}

\begin{figure}
\begin{center}
\includegraphics[width=13cm]{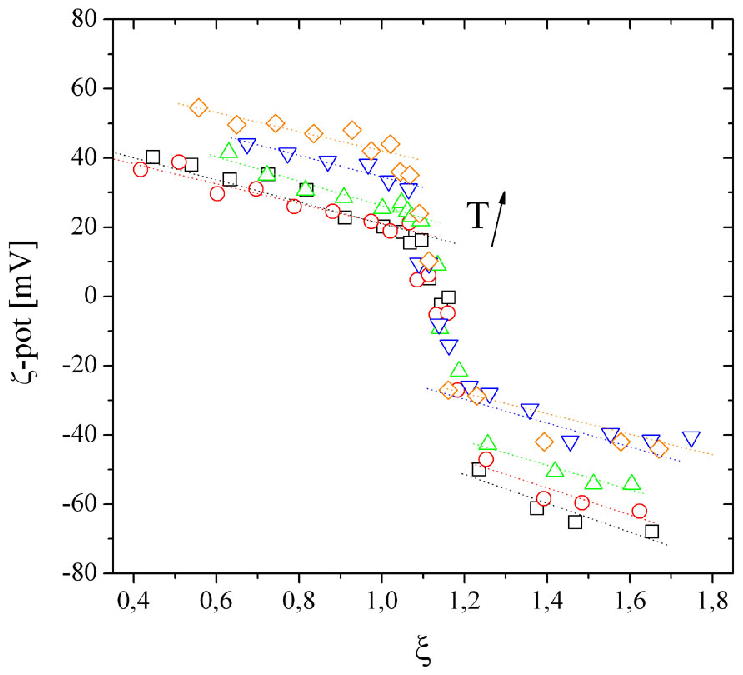}
\vspace{1cm} \caption{\label{fig:ZT}}
\end{center}
\end{figure}

Fig. \ref{fig:ZT} - $\zeta$-potential curves as a function of the
charge ratio $\xi$, for the different investigated temperatures
($\square$:  5 $^{\circ}$C, ${\circ}$:  20 $^{\circ}$C,
$\bigtriangleup$: 40 $^{\circ}$C,
 $\bigtriangledown$: 60 $^{\circ}$C,  $\diamond$: 80 $^{\circ}$C). Lines
 represent linear fit on data, in regions below and above the
 isoelectric condition.

\newpage

\clearpage

\clearpage

\clearpage

\clearpage

\clearpage

\newpage
\newpage

\end{document}